\documentclass[10pt,journal]{article}
\usepackage[utf8]{inputenc}
\usepackage{amsmath,amssymb,amsthm}
\usepackage{graphicx}
\usepackage{float}
\usepackage{booktabs}
\usepackage{geometry}
\usepackage{siunitx}
\usepackage{xcolor}
\usepackage[hidelinks]{hyperref}
\usepackage{parskip}
\newtheorem{remark}{Remark}

\title{\textbf{Monotonic, Minimum-Settling-Time PI Tuning for\\First-Order-Plus-Dead-Time Plants:\\A Tangency Characterization}}
\author{\c{S}enol G\"ulg\"on\"ul\\
Department of Electrical and Electronics Engineering\\
Ostim Technical University, Ankara, Turkey\\
\texttt{senol.gulgonul@ostimteknik.edu.tr}}
\date{}

\begin{document}
\maketitle

\begin{abstract}
This paper studies PI tuning of a first-order-plus-dead-time (FOTD) plant for the
fastest strictly monotone (zero-overshoot) setpoint step response, with monotonicity
imposed on the plant output only. The minimizer is shown to be neither the pole-zero
cancellation design nor the multiple real dominant pole (MRDP) design. It is a
non-cancellation point at which the closed loop carries a slow real mode of small
residue together with a faster underdamped complex pair, with the controller zero placed
near the dominant real pole. The analytical centerpiece is a single tangency identity,
$\tan(\omega\tau^\star+\alpha)=(a-b)/\omega$, which states that the monotonicity boundary
is the locus where the secondary complex mode just fails to drive the output slope below
zero. From this identity the design reduces to nested scalar conditions, realized at three
levels of fidelity: an explicit closed-form rule, an exact response-based reduction, and a
simulation-free transcendental system whose only non-elementary step is a fourth-order
polynomial root. Relative to the critically damped Lambert-W cancellation rule the design
reduces the 2\% settling time by 14 to 52 percent and lowers the load integrated absolute
error by 5 to 38 percent. We report the full cost: for delay-dominated plants ($T/L\lesssim0.55$) the
control stays one-pulse, so the design is itself admissible in Huba's sense and merely faster, but for
larger lag ratios the control becomes two-pulse, and across the range the maximum sensitivity rises
from 1.39 to between 1.44 and 1.62.
The contribution is therefore positioned not as a uniformly better tuning but as the exact
characterization of a specific, well-defined operating point, together with an honest
multi-metric comparison against established rules.
\end{abstract}

\section{Introduction}

The PI controller remains the workhorse of process control, and tuning rules for the
first-order-plus-dead-time (FOTD) model
\begin{equation}
G(s)=\frac{K\,e^{-Ls}}{Ts+1}
\label{eq:plant}
\end{equation}
fill handbooks \cite{odwyer}. Among the many objectives, a strictly monotone
(non-overshooting) setpoint response is required wherever overshoot is unacceptable, for
example in mechatronic positioning, thermal processes, and any loop feeding a downstream
stage that must not be driven past its target. Within the monotone class, the practically
relevant secondary objective is speed, that is, the smallest settling time.

Three reference designs frame the problem. The Chien-Hrones-Reswick rules \cite{chr}, recently given
an analytical solution in \cite{gulgonul1}, target
zero or small overshoot empirically. The pole-zero cancellation design, in which the
controller zero cancels the plant pole and the loop reduces to a scalar delay equation
solvable by the Lambert W function \cite{corless}, gives a clean monotone response but is
conservative in speed and, as noted by \r{A}str\"om and H\"agglund \cite{astrom} and by Skogestad
\cite{skogestad}, leaves the cancelled lag in the load response. The multiple real dominant pole (MRDP) method used by
Huba and co-workers \cite{huba2013,huba2016,huba2018,huba2021} places a triple real pole and is, under the
shape constraints of that framework, close to the fastest monotone design. That framework is
deliberately broad, and its monotone-output ideal is paired with a control-shape requirement.
Theorem~1 of \cite{huba2013} establishes, for integral and unstable first-order plants, that a
monotone output corresponds to a one-pulse control, an input made of two monotone segments
separated by a single extremum, which is appropriate for amplitude-limited actuators such as
pulse-width-modulated drives \cite{huba2024}. Deviations from the monotone and one-pulse ideals are quantified by
the $TV_0$ and $TV_1$ measures, and the optimal tuning is obtained numerically by a performance
portrait, the only closed-form anchor being the triple real pole, which the same authors note is
more conservative than the numerical optimum. Two further facts from that line of work bear directly
on the present problem. First, the FOTD plant has been studied in the same framework \cite{huba2016},
and a two-pulse class has been characterized for second-order and double-integrator plants
\cite{hubabelai2014}, where a monotone output of a second-order plant inverts to a two-pulse input
measured by $TV_2$. Second, and central here, Remark~2 of \cite{huba2013} notes that for a first-order
plant whose dead time is realized through a dynamical term rather than as a pure time shift, the
optimal input becomes an $nP$ function with $n>1$. The two-pulse control reported below is precisely
this case, the $n=2$ realization for a first-order plant, and it is therefore distinct in origin from
the second-order two-pulse class of \cite{hubabelai2014}: here the second pulse is induced by the
dead time, not by a second plant integrator. In the comparisons below, the curves labeled MRDP are
this raw triple-real-pole tuning, which satisfies the one-pulse, monotone-output admissibility only
for $T/L\lesssim1.5$; where it is plotted beyond that range, as at $T/L=5$ in Figure~\ref{fig:pulse},
it has itself left the admissible set, becoming two-pulse and overshooting, and is shown to expose
that limit rather than as a design the framework would deploy.

This paper isolates a narrower and sharply defined problem: minimize the 2\% settling time
subject to monotonicity of the output, with the control signal left unconstrained. We show
that the optimizer of this problem departs from both cancellation and MRDP, characterize it
by a tangency identity, reduce it to scalar conditions, and compare it fairly against the
reference rules on a multi-metric basis that reports the costs as well as the gains.

\section{Problem formulation}

The controller is PI,
\begin{equation}
C(s)=K_p+\frac{K_i}{s}=K_p\,\frac{s+z}{s},\qquad z=\frac{K_i}{K_p}.
\end{equation}
We normalize $K=1$ and $L=1$, so the loop depends on the single shape parameter $T/L$; gains
de-normalize as $K_{p,\mathrm{real}}=K_p/K$ and $K_{i,\mathrm{real}}=K_i/(KL)$. The
closed-loop characteristic quasipolynomial is
\begin{equation}
\Delta(s)=T s^2+s+(K_p s+K_i)\,e^{-s}=0,
\label{eq:char}
\end{equation}
and the setpoint step response, for $t>L$, is the modal sum over the roots $p_k$ of
\eqref{eq:char},
\begin{equation}
y(t)=1+\sum_k c_k\,e^{p_k(t-1)},\qquad
c_k=\frac{K_p(p_k+z)}{p_k\,\Delta'(p_k)}.
\label{eq:modal}
\end{equation}
The controller zero is not a separate mode; it enters every residue $c_k$ through the
numerator factor $(p_k+z)$, since a zero adds no pole. A zero placed near a pole therefore
shrinks that mode's residue, and at the dominant real pole $p_1=-a$ one has
$c_a\propto(z-a)$, which vanishes as $z\to a$. This residue reshaping, rather than any new
term, is the entire effect of the zero and is the mechanism exploited below.
The design problem is
\begin{equation}
\min_{K_p,K_i}\;T_s\quad\text{s.t.}\quad y'(t)\ge 0\ \ \forall t\ge 0,
\label{eq:problem}
\end{equation}
where $T_s$ is the 2\% settling time and $y'(t)\ge 0$ is output monotonicity.

\section{Coordinates and modal structure}

We use the pole-zero coordinates $(a,z)$, where $a$ is the magnitude of the dominant real
closed-loop pole and $z$ is the controller-zero magnitude. Imposing $\Delta(-a)=0$ in
\eqref{eq:char} yields the exact gain map
\begin{equation}
K_p=\frac{a(1-Ta)}{(z-a)\,e^{a}},\qquad K_i=z K_p,
\label{eq:map}
\end{equation}
valid for $0<a<1/T$ and $z>a$. The map fixes the real pole at $-a$; the remaining spectrum,
in particular the dominant complex pair $p_c=-b+j\omega$, is determined by $(a,z)$ through
\eqref{eq:char}. Retaining the two dominant modes, the step response and its slope are
\begin{align}
y(\tau)  &= 1 + c_a e^{-a\tau} + 2|c_c| e^{-b\tau}\cos(\omega\tau+\phi),\\
y'(\tau) &= A e^{-a\tau} + D e^{-b\tau}\cos(\omega\tau+\alpha),
\label{eq:twomode}
\end{align}
with $\tau=t-1$, $A=-a\,c_a$, $D=2|c_c p_c|$, $\alpha=\arg(c_c p_c)$, $\phi=\arg(c_c)$. At
the optimum the real residue $c_a$ is negative and small, because the zero sits near the
real pole, so the slow real mode is a low-amplitude correction whose sole role is to keep
the faster complex pair from breaking monotonicity. The structure is irreducibly three-modal:
a slow real pole, a fast complex pair, and the shared zero.

\section{The tangency identity}

Output monotonicity is non-negativity of the impulse response $h(t)=y'(t)$. Its boundary is
the locus where the slope touches zero without crossing, that is, a double zero of $y'$:
\begin{equation}
y'(\tau^\star)=0,\qquad y''(\tau^\star)=0.
\label{eq:dbl}
\end{equation}
Differentiating \eqref{eq:twomode},
\begin{equation*}
y''(\tau)=-aA e^{-a\tau}-D e^{-b\tau}\big[b\cos(\omega\tau+\alpha)+\omega\sin(\omega\tau+\alpha)\big].
\end{equation*}
Writing $\theta=\omega\tau^\star+\alpha$, the first condition in \eqref{eq:dbl} gives
$A e^{-a\tau^\star}=-D e^{-b\tau^\star}\cos\theta$. Substituting into the second and cancelling
the common positive factor $D e^{-b\tau^\star}$ leaves $(a-b)\cos\theta-\omega\sin\theta=0$,
that is,
\begin{equation}
\boxed{\;\tan(\omega\tau^\star+\alpha)=\dfrac{a-b}{\omega}.\;}
\label{eq:tangency}
\end{equation}
This tangency identity is the main analytical result. It states that at the monotonicity
boundary the phase of the binding dip is fixed by the pole geometry alone, through the real-pole
magnitude $a$, the complex-pair real part $b$, and the imaginary part $\omega$. Because $A,D>0$
the balance requires $\cos\theta<0$, while $a<b$ makes the right side of \eqref{eq:tangency}
negative so $\sin\theta>0$; the relevant root therefore lies in the second quadrant, and the
dip time is explicit,
\begin{equation}
\tau^\star=\frac{\pi+\arctan\!\big((a-b)/\omega\big)-\alpha}{\omega}.
\label{eq:taustar}
\end{equation}
Numerically \eqref{eq:tangency} and \eqref{eq:taustar} are exact at the optima: for $T/L=1$ the
formula gives $\tau^\star=4.71$, matching the dip located by full simulation, and for $T/L=3$ the
two sides of \eqref{eq:tangency} read $-0.987$ and $-0.978$. The binding dip is late and lies
inside the 2\% band, so the deeper poles have decayed before the constraint binds, which is why
the two-mode reduction is faithful.

\begin{figure}[H]
\centering
\includegraphics[width=0.80\linewidth]{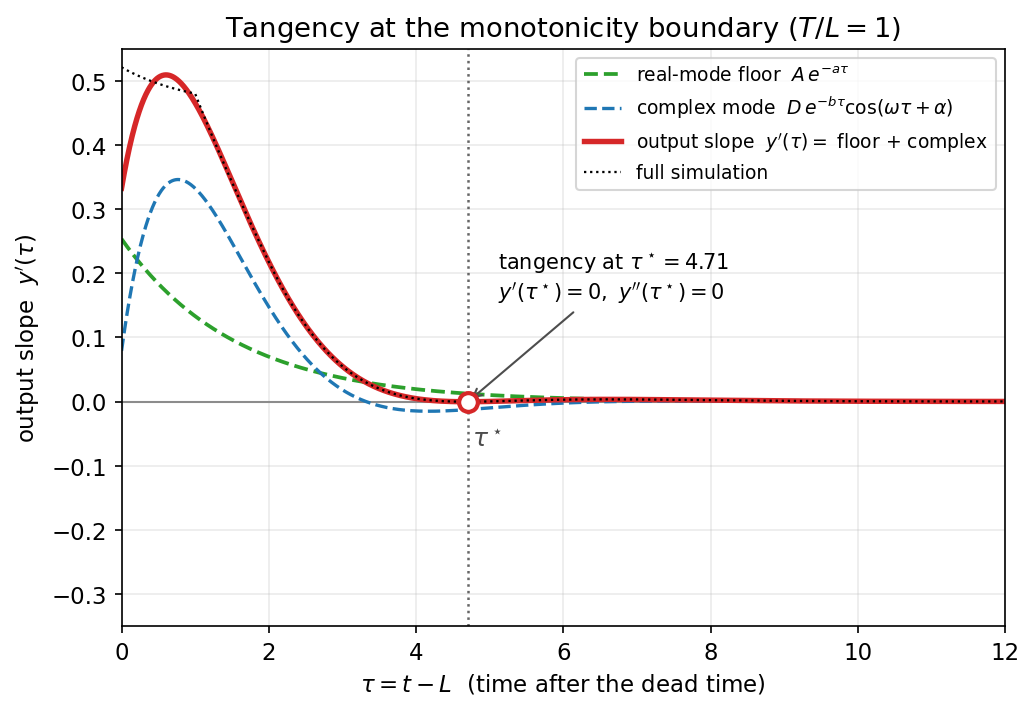}
\caption{Tangency mechanism at $T/L=1$: the output slope $y'(\tau)$ (red) is the sum of the positive
real-mode floor (green) and the decaying complex mode (blue), and it touches zero tangentially at
$\tau^\star=4.71$, where $y'(\tau^\star)=y''(\tau^\star)=0$; the two-mode reconstruction matches the
full simulation (dotted).}
\label{fig:tangency}
\end{figure}

Figure~\ref{fig:tangency} shows the mechanism at $T/L=1$. The output slope is the sum of a positive
real-mode floor $A\,e^{-a\tau}$ and a decaying oscillation $D\,e^{-b\tau}\cos(\omega\tau+\alpha)$.
Near $\tau^\star$ the oscillation is in its first negative swing, and the floor lifts it so that the
slope grazes zero rather than crossing it: the amplitude balance $y'(\tau^\star)=0$ makes the two
modes cancel, and the horizontal tangent $y''(\tau^\star)=0$ makes the contact a graze. Eliminating
the amplitudes $A$ and $D$ between those two conditions leaves the phase relation~\eqref{eq:tangency},
so the binding instant and its phase are fixed by the pole geometry alone, independently of the mode
amplitudes.

\begin{remark}
The same construction at a triple real point recovers MRDP, $\Delta=\Delta'=\Delta''=0$. The
present optimum is not such a multiplicity: tested on the optima, $|\Delta'(p_c)|$ ranges from
0.8 to 4.7, so the pair is not a repeated root, and $y'''(\tau^\star)>0$, so the contact is a
simple convex tangency, not a degenerate one. MRDP is the critically damped $\omega\to0$ limit;
the present design is its underdamped continuation into the interior, governed by
\eqref{eq:tangency} rather than by a multiplicity.
\end{remark}

\section{Reduction and solvers}

Equation \eqref{eq:tangency} with the amplitude balance $y'(\tau^\star)=0$ fixes the zero
$z=z(a)$ on the monotonicity boundary; settling stationarity then selects $a$. The two-dimensional
search thus collapses to nested scalar conditions, which we realize at three levels.

\begin{figure}[H]
\centering
\includegraphics[width=0.74\linewidth]{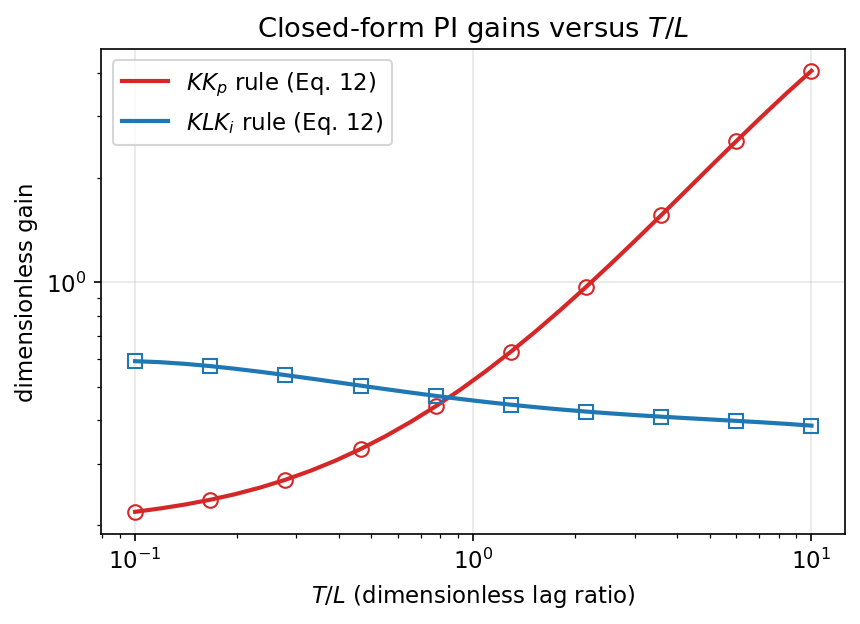}
\caption{Dimensionless PI gain groups $KK_p$ and $KLK_i$ of Eq.~\eqref{eq:quartic} versus the lag
ratio $T/L$, with the numerically exact optima as open markers.}
\label{fig:gains}
\end{figure}

\emph{Level 1 (explicit rule).} Fitting the located optimum directly in the gains, with
$u=\ln(T/L)$,
\begin{align}
K_p &= \frac{1}{K}\,\exp\!\big(-0.0060u^4-0.0138u^3+0.1432u^2+0.7079u-0.6518\big),\nonumber\\
K_i &= \frac{1}{KL}\,\exp\!\big(-0.0027u^4+0.0041u^3+0.0232u^2-0.1145u-0.7849\big).
\label{eq:quartic}
\end{align}
Over $T/L\in[0.1,10]$ this rule holds settling within $0.10L$ of the optimum with residual overshoot
below 0.05\%, and reproduces the exact optimal gains to within 1.6\% in $K_p$ and 0.3\% in $K_i$, as
Figure~\ref{fig:gains} shows. The design is scale-invariant: with time measured in units of $L$ the
loop depends only on $T/L$, so the dimensionless groups $KK_p$ and $KLK_i$ are functions of $T/L$
alone, for any $K$ and $L$. The proportional group rises roughly linearly with the lag ratio while the
integral group falls gently.

\emph{Level 2 (response-based).} For trial $a$, an inner scalar solve drives the true response onto
the monotone edge to obtain $z(a)$; the outer scalar solve minimizes the 2\% settling. No fit is
used; the gains match the brute-force optimum with settling within $0.018L$.

\emph{Level 3 (simulation-free).} The inner check is replaced by \eqref{eq:tangency} and the
amplitude balance on the dominant pair obtained from \eqref{eq:char}. With a Pade(2,2) surrogate of
the delay the pair becomes the dominant root of a quartic, closed form in radicals; over $T/L\ge0.3$
the gains agree with the optimum to 1 to 2 percent and the full response stays monotone. The single
non-elementary operation is the quartic root.

\begin{remark}
The optimum is transcendental: even with the poles in hand the monotone boundary and the settling
instant are transcendental in time. A fully elementary closed form does not exist, exactly as the
cancellation rule is closed form only through the Lambert W function. The tangency identity is the
analytic object; the final scalar conditions are solved numerically.
\end{remark}

\section{Structural results}

\paragraph{Non-cancellation.} The minimizer does not satisfy $K_p=T K_i$. The ratio $K_p/(TK_i)$
runs from 1.05 to 3.67 across $T/L$, never unity, so the optimum lies strictly off the cancellation
line. The secondary pair is underdamped, with damping $\zeta$ from 0.65 to 0.89, below $\sqrt3/2$
for almost the whole range. The minimum-settling solution is therefore not the cancellation design,
which refines the all-pole intuition that minimum settling requires cancellation: the dead time
supplies a faster complex mode that cancellation would suppress.

\paragraph{Two regimes.} Rescaling all positions so $b=1$, the dominant real pole is always the
rightmost (slowest) pole, but the zero moves with $T/L$. For delay-dominated plants the order is
zero $<$ pair $<$ pole, with the real pole carrying a large residue and the zero shaping the pair
from the left. Near $T/L=1$ the zero crosses the pair, and for lag-dominated plants the order becomes
pair $<$ zero $<$ pole, with the zero collapsing onto the real pole, that is, cancellation. The
residue hands off from the real pole (residue 0.85 down to 0.02) to the complex pair (residue 0.26
up to 0.81) across this transition. The two ends are analytical anchors: a delay-free third-order
model at one end and the Lambert W cancellation rule at the other.

\section{Performance and comparison}

All comparisons use the cancellation (Lambert W) rule with $K_i=1/(KeL)$, $K_p=TK_i$, which holds a
settling time of $6.53L$ and a maximum sensitivity of 1.39 for every $T/L$.

\begin{table}[H]
\centering
\caption{Settling time $T_s/L$ versus $T/L$ for the cancellation rule, the MRDP tuning, and the
proposed design.}
\label{tab:ts}
\begin{tabular}{cccc}
\toprule
$T/L$ & Lambert W ($\gamma=1$) & MRDP (triple) & This work \\
\midrule
0.1 & 6.53 & 4.43 (mono) & 3.15 \\
1.0 & 6.53 & 6.53 (mono) & 4.34 \\
10  & 6.53 & 12.87 (20.9\% OS) & 5.62 \\
\bottomrule
\end{tabular}
\end{table}

\begin{table}[H]
\centering
\caption{Multi-metric comparison of the three tunings, where OS is output overshoot, ``shape'' is the
control-signal shape (mono/1P/2P = monotone/one-pulse/two-pulse), $M_s$ is maximum sensitivity, and
$\mathrm{IAE}_d=1/K_i$ is the load integrated absolute error.}
\label{tab:multi}
\begin{tabular}{llccccc}
\toprule
$T/L$ & design & $T_s/L$ & OS & shape & $M_s$ & $\mathrm{IAE}_d$ \\
\midrule
0.1 & Lambert W   & 6.53 & 0\%    & mono & 1.39 & 2.72 \\
0.1 & MRDP (Huba) & 4.43 & 0\%    & mono & 1.49 & 2.03 \\
0.1 & This work   & 3.15 & 0\%    & mono & 1.62 & 1.69 \\
\midrule
1.0 & Lambert W   & 6.53 & 0\%    & mono & 1.39 & 2.72 \\
1.0 & MRDP (Huba) & 6.53 & 0\%    & mono & 1.39 & 2.72 \\
1.0 & This work   & 4.34 & 0\%    & 2P   & 1.54 & 2.19 \\
\midrule
10  & Lambert W   & 6.53 & 0\%    & 1P   & 1.39 & 2.72 \\
10  & MRDP (Huba) & 12.87 & 20.9\% & 2P  & 1.60 & 1.01 \\
10  & This work   & 5.62 & 0\%    & 2P   & 1.44 & 2.59 \\
\bottomrule
\end{tabular}
\end{table}

\begin{figure}[H]
\centering
\includegraphics[width=0.82\linewidth]{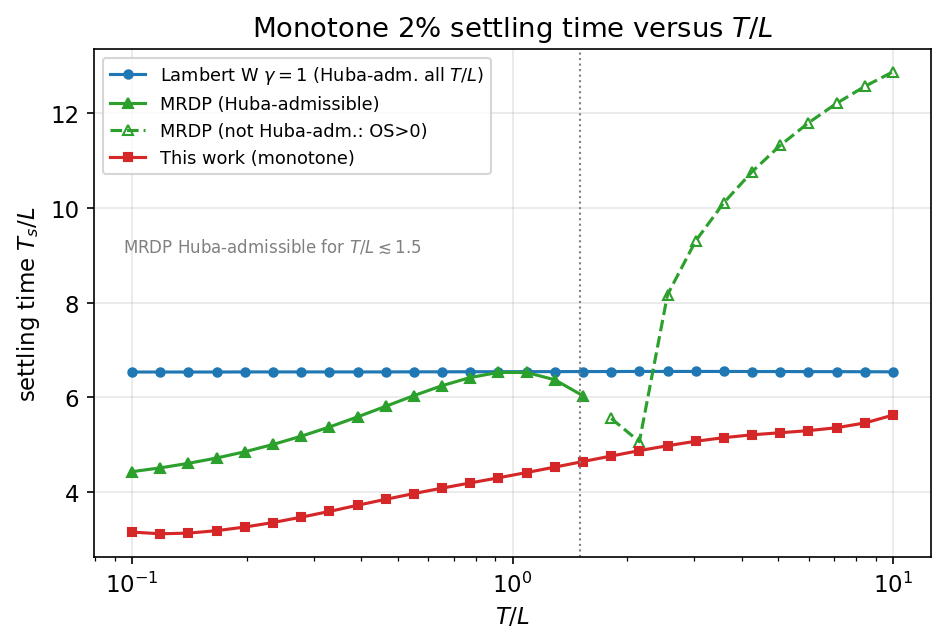}
\caption{Monotone 2\% settling time versus $T/L$ for the cancellation rule, the MRDP (Huba) tuning
(solid where Huba-admissible, dashed where not), and the proposed design.}
\label{fig:ts}
\end{figure}

Tables~\ref{tab:ts} and \ref{tab:multi} give the honest picture. The proposed design reduces the
2\% settling time by 14 to 52 percent across $T/L$ (largest for delay-dominated plants) and lowers the
load $\mathrm{IAE}_d$ by 5 to 38 percent, since it runs a larger integral gain. The settling gain is
plotted in Figure~\ref{fig:ts}, together with the MRDP (Huba) tuning. A point on the latter deserves
care: MRDP is a valid Huba design, meaning monotone output \emph{and} one-pulse control, only for
$T/L\lesssim1.5$. We confirmed the boundaries numerically. The MRDP output overshoot is below 0.05
percent up to $T/L\approx1.6$, reaches 0.83 percent at $T/L=2$, and grows to 4.6 percent at $T/L=3$,
11.8 percent at $T/L=5$, and 20.9 percent at $T/L=10$; its control stays one-pulse only up to
$T/L\approx2.5$ and is two-pulse beyond. For delay-dominated plants MRDP is admissible and faster than
the cancellation rule, coinciding with it at $T/L=1$. For $T/L>1$ it leaves the admissible set, so out
there it is no longer Huba's deployed design but only the raw analytic anchor; Huba's
performance-portrait method would instead back off to a slower one-pulse monotone design. The
cancellation rule, by contrast, stays Huba-admissible for every $T/L$ and is the proper one-pulse
reference. The MRDP settling curve is itself revealing: its small overshoot stays inside the 2\% band
up to $T/L\approx2.4$, so settling even dips below the cancellation value near $T/L=2$, but once the
overshoot exceeds the band the settling rises steeply, reaching $12.9L$ at $T/L=10$. The proposed
design stays monotone and fastest throughout, at the price of a two-pulse control.

\begin{figure}[H]
\centering
\includegraphics[width=0.72\linewidth]{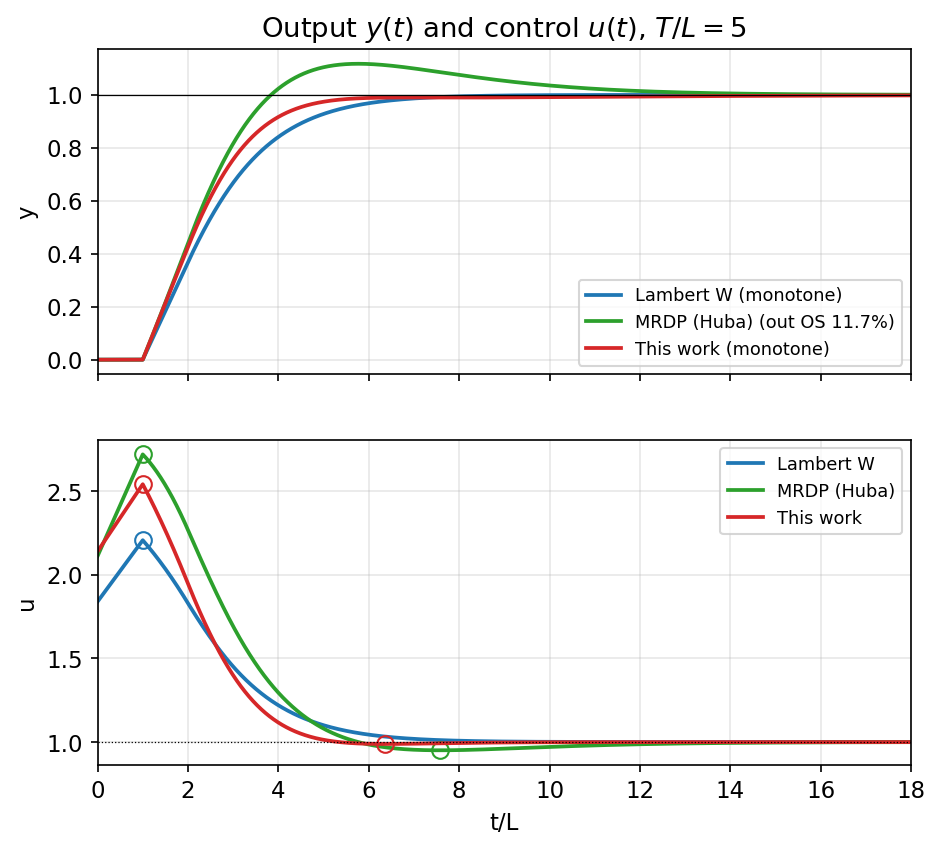}
\caption{Output (top) and control (bottom) at $T/L=5$ for the cancellation, the MRDP (Huba) tuning,
and the proposed design, with control extrema circled and the step jump shown at $t=0$.}
\label{fig:pulse}
\end{figure}

The speed is paid for in the control signal, but only beyond a sharp threshold. For $T/L\lesssim0.55$
the proposed control is exactly monotone, rising to the steady value without overshoot; at
$T/L\approx0.56$ it turns two-pulse, a slight overshoot of the steady value appearing and then a dip
of about 1.3 percent below it, and from there the overshoot grows (6 percent at $T/L=1$, 17 percent at
$T/L=1.5$) while the dip slowly shrinks. The two-pulse control carries two extrema and lies outside the
one-pulse admissible set of the Huba framework. The monotone regime has a consequence worth stating:
for delay-dominated plants the proposed design is itself Huba-admissible, with monotone output and
one-pulse (monotone) control, and it is faster there than the MRDP tuning, which is also admissible,
settling in $3.15L$ against $4.43L$ at $T/L=0.1$ and $3.74L$ against $5.90L$ at $T/L=0.5$. In that
regime we do not relax Huba's constraint at all but improve on MRDP inside it, the trade being
robustness rather than control shape, since the proposed $M_s$ is higher there ($1.62$ against MRDP's
$1.49$ at $T/L=0.1$). As the lag ratio grows the initial proportional action and the control peak grow
with it, since any design must push harder on a slower plant. Figure~\ref{fig:pulse} shows the contrast
at $T/L=5$, where the
divergence is sharpest. The cancellation control remains one-pulse and its output monotone; the
proposed design keeps a monotone output with a two-pulse control; and the raw MRDP tuning has lost
both properties, overshooting the output by 11.8 percent and itself becoming two-pulse. This mirrors
the settling spike of Figure~\ref{fig:ts}: the triple-real-pole design degrades on every axis for
strongly lag-dominated plants, while the proposed design stays monotone and fastest.

\begin{remark}
The control-shape transition and the output-monotonicity boundary are two distinct thresholds. The
control turns two-pulse at $T/L\approx0.56$, whereas the output remains strictly monotone, with output
total variation $TV_0$ at the numerical floor, up to $T/L\approx0.66$. For larger lag ratios the
explicit-rule optima carry a residual $TV_0$ below $10^{-4}$, which can be driven to zero by enforcing
strict monotonicity at under 5 percent additional settling; the control stays two-pulse in either case,
so the two-pulse shape is not a consequence of that residual. This is consistent with the Huba
framework rather than a departure from it. Theorem~1 of \cite{huba2013} bounds the ideal control of a
first-order plant, with the dead time treated as a pure time shift, to one pulse; Remark~2 of the same
work predicts that once the dead time enters the loop as a dynamical element the input becomes $nP$
with $n>1$. The present two-pulse control is that $n=2$ instance for a first-order plant, induced by
the delay rather than by a second plant integrator, and hence distinct from the second-order two-pulse
class of \cite{hubabelai2014}.
\end{remark}

\begin{figure}[H]
\centering
\includegraphics[width=0.62\linewidth]{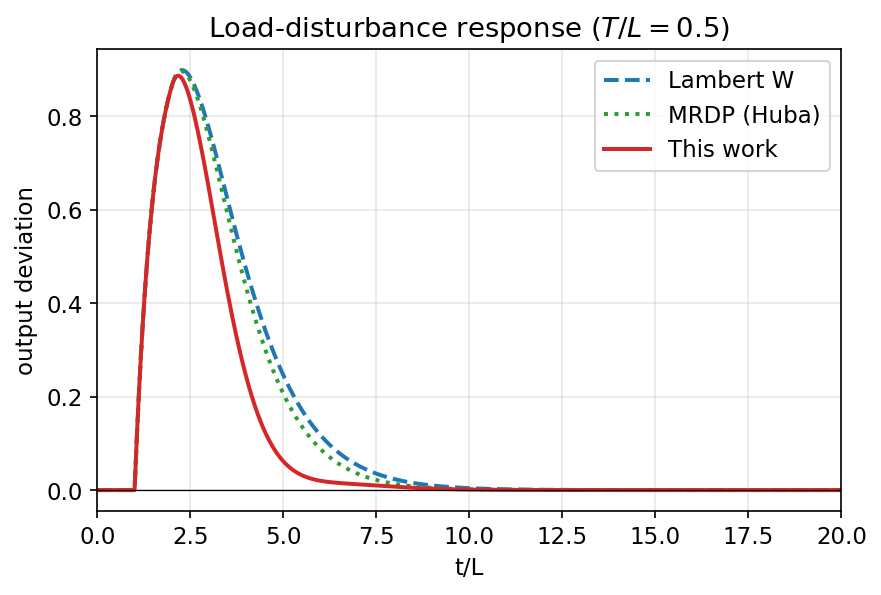}
\caption{Load-disturbance response at $T/L=0.5$ for the cancellation rule, the MRDP (Huba) tuning, and
the proposed design.}
\label{fig:loadresp}
\end{figure}

The load-disturbance rejection is shown in Figure~\ref{fig:loadresp} for $T/L=0.5$. A unit step load
enters at the plant input and is rejected back to zero by the integral action. The three tunings
produce a similar peak deviation, near 0.89, but the proposed design returns to zero markedly faster
because of its larger integral gain, giving a smaller integrated error; the MRDP and cancellation
responses are close, with MRDP recovering slightly faster than the cancellation rule.

\begin{figure}[H]
\centering
\includegraphics[width=0.62\linewidth]{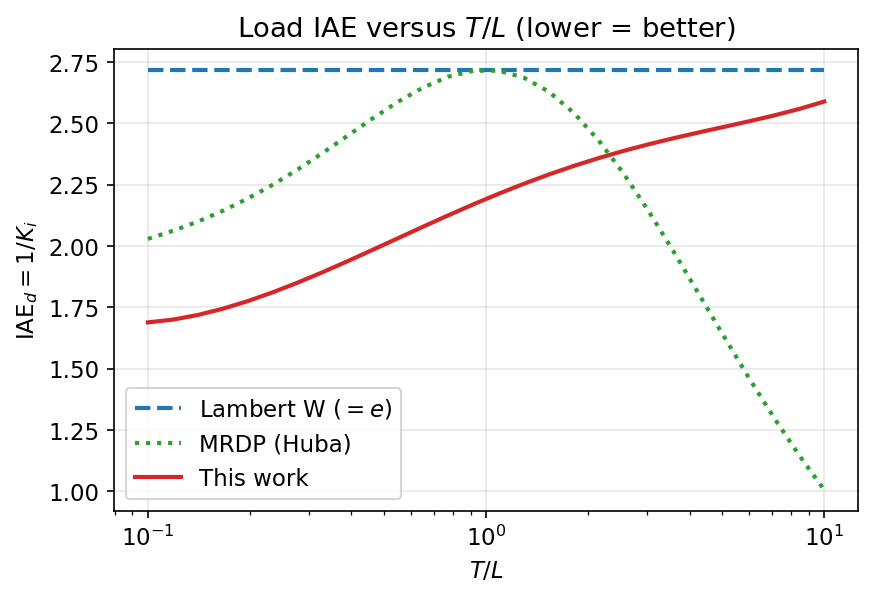}
\caption{Load IAE $\mathrm{IAE}_d=1/K_i$ versus $T/L$ for the cancellation rule, the MRDP (Huba)
tuning, and the proposed design.}
\label{fig:loadiae}
\end{figure}

The integrated cost across the lag range is summarized in Figure~\ref{fig:loadiae}. For any stabilizing
PI loop the integrated absolute error after a unit input load step equals $1/K_i$ exactly, an
identity already noted in \cite{huba2013}, while the
deviation keeps one sign, so the curve is simply $1/K_i$ and a larger integral gain rejects load
better. The proposed design runs 5 to 38 percent below the flat cancellation value $e$ across $T/L$,
the largest margin for delay-dominated plants. MRDP undercuts even the proposed design for strongly
lag-dominated plants, but only by overshooting the output, so that gain is not admissible.

\begin{figure}[H]
\centering
\includegraphics[width=0.62\linewidth]{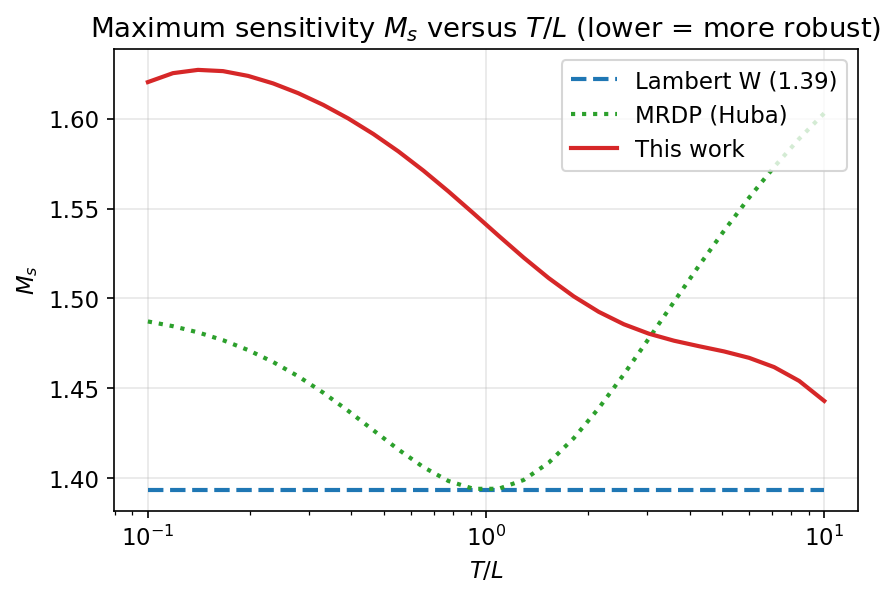}
\caption{Maximum sensitivity $M_s$ versus $T/L$ for the cancellation rule, the MRDP (Huba) tuning, and
the proposed design.}
\label{fig:ms}
\end{figure}

The price of the added speed and load rejection is robustness, Figure~\ref{fig:ms}. The proposed
design raises the maximum sensitivity from the 1.39 of the cancellation rule to between 1.44 and 1.62,
with the largest penalty exactly where the speed gain is largest, for delay-dominated plants, while
the complementary sensitivity peak stays at 1.0, consistent with the monotone output. MRDP matches the
cancellation robustness for delay-dominated plants but its $M_s$ climbs above the proposed design
beyond $T/L\approx3$, so no single tuning is uniformly most robust.

\section{Positioning}

The framework within which this problem sits is due to Huba and co-workers and is prior art: the
monotone-output pulse-count taxonomy, with one-pulse control for integral and first-order plants
\cite{huba2013} and two-pulse control for second-order and double-integrator plants
\cite{hubabelai2014}; the $TV_0$, $TV_1$ and $TV_2$ shape measures; the performance-portrait
numerical optimum; the triple real pole analytic rule and its acknowledged suboptimality; the FOTD
performance-limit study \cite{huba2016,huba2018}; and the load identity $\mathrm{IAE}_d=1/K_i$
\cite{huba2013}. The objective of a fastest monotone, non-overshooting response is itself classical
and is not claimed here as new; the present work is the closed-form characterization of that optimum
for a specific plant. It completes a closed-form monotonic minimum-settling program developed by the
author for all-pole plants up to third order \cite{gulgonul2} and for the pure-delay plant
\cite{gulgonul3}: the FOTD plant is the lag-plus-delay member, recovering the all-pole treatment as
$L\to0$ and the pure-delay boundary-contact characterization as $T\to0$. The present contribution is
specific and complementary to that body of work. First,
under output-only monotonicity, with the control left unconstrained rather than jointly constrained
to a one-pulse shape as in \cite{huba2013}, the cancellation and MRDP designs are not optimal, and the
optimum is a non-cancellation, zero-near-pole design that is faster and rejects load better. Second,
the optimum is characterized in closed form by the tangency identity \eqref{eq:tangency}, which makes
the monotonicity boundary explicit and reduces the search to nested scalar conditions at three levels
of fidelity; it supplies for the stable FOTD plant the analytic object that the triple real pole only
approximates and that the performance portrait otherwise reaches only numerically. Third, the trade is quantified end to end, and it is
regime-dependent. For delay-dominated plants ($T/L\lesssim0.55$) the proposed control is itself
one-pulse, so the design is Huba-admissible and faster than MRDP, the only cost being a higher $M_s$.
For larger lag ratios the control becomes two-pulse and leaves the admissible set, and the speed and
load-rejection gains are then paid for with both the two-pulse control and the higher $M_s$. We do not
claim a uniformly superior tuning. The proposed design is faster and lighter on load IAE; for balanced
and lag-dominated plants the Huba design respects a one-pulse input and a lower $M_s$; the cancellation
rule is the most robust and has the cleanest control. Each is
Pareto-optimal for its own constraint set, and the present work supplies the missing analytical
point of that front together with its exact characterization.

\section{Conclusion}

The output-monotone minimum-settling PI tuning of an FOTD plant is governed by a single tangency
identity that pins the monotonicity boundary to the closed-loop pole geometry. From it the design
reduces to nested scalar conditions, realized as an explicit rule, an exact response-based reduction,
and a simulation-free transcendental system. The design is faster than the cancellation and MRDP
rules and rejects load disturbances better, with the cost appearing as a two-pulse control and a
modest loss of robustness rather than as output overshoot. The optimum is transcendental; the
tangency identity is the analytic object that organizes it.

\end{document}